# Ultrafast Large Area Micropattern Generation in Non-absorbing Polymer Thin-Films by Pulsed Laser Diffraction

*Ankur Verma, Ashutosh Sharma\* and Giridhar U. Kulkarni*

[*]     Prof. A. Sharma, A. Verma
Department of Chemical Engineering and DST Unit on Nanoscience, Indian Institute of Technology Kanpur, Kanpur, 208016 (India)
E-mail: ashutos@iitk.ac.in
        Prof. G.U. Kulkarni
Chemistry and Physics of Materials Unit and DST Unit on Nanoscience, Jawaharlal Nehru Centre for Advanced Scientific Research, Bangalore, 560 064 (India)

We report an ultrafast, parallel and beyond-the-master micro-patterning technique for ultrathin (30 nm–400 nm) non-absorbing polymer films by diffraction of a laser light through a two dimensional periodic aperture. The redistribution of laser energy absorbed by the substrate causes self-organization of polymer thin-film in the form of wrinkle like surface relief structures caused by localized melting and freezing of the thin-film. Unlike the conventional laser ablation and laser writing processes, low laser fluence is employed to only passively swell the polymer as a pre-ablative process without the loss of material, and without absorption/reaction with the incident radiation. Self-organization in the thin polymer film aided by the diffraction pattern produces micro-structures made up of thin raised lines. These regular microstructures have far more complex morphologies than the mask geometry and very narrow line widths that can be an order of magnitude smaller than the openings in the mask. The microstructure morphology is easily modulated by changing the film thickness, aperture size and geometry and by changing the diffraction pattern, e.g., by changing the aperture-substrate distance.

# 1. Introduction

Micropattern generation[1,2] by the multi-step top-down techniques such as photolithography[3] and by the self-organization based methods such as controlled dewetting,[4,5] electric field lithography,[6] etc. produce the pattern morphology and the pattern length scales that are commensurate with the mask, template or the master employed. We present here an ultra-fast, single step technique of beyond-the-mask pattern generation in non-absorbing thin (< 500 nm) polymer films by harnessing the self-organization induced by the low fluence laser diffraction patterns. In contrast to the nature of patterns generated by other techniques, the micro-patterns obtained here by simple masks are in the form of complex lattice-structures with line widths that can be tuned by the film thickness and can be more than an order of magnitude smaller than the mask openings.

The fabrication of patterned and functionalized polymeric surfaces is an area of increasing interest because of potential applications in microfluidics, microelectromechanical systems (MEMS), polymer electronics, multifunctional coatings and bioengineering.[1] In addition, patterned polymeric surfaces provide an inexpensive route to pattern other kinds of surfaces such as metals and ceramics.[2] Various lithography techniques are used to produce surface structures on polymer films and substrates.[3] The most widely used UV lithography requires light sensitive polymers (photoresists) and involves multi step fabrication protocol (mask design, exposure, pattern developing etc.). Lithography techniques produce the pattern geometry, pattern length scales and line widths that are commensurate with the mask openings.

There are also several techniques based on self-organization in soft materials for generation of controlled micro-patterns that can be contrasted with the thermal energy induced self-organization studied here. For example, controlled dewetting of ultrathin (<100 nm) polymer films have been extensively studied as a promising tool for polymer patterning. Polymer thin-films on physico-chemically patterned substrates produce patterned array of polymer droplets, when destabilized by either heating above glass transition temperature ($T_g$) or exposure to solvent vapor.[4] The characteristic length scales (feature size and wavelength) of dewetting strongly depend on the initial thickness of polymer film.[5] The time scales involved in dewetting of highly viscous polymer films are typically of the order of several minutes and the structures can be aligned on the scale of physico-chemically patterned templates used as substrates for dewetting. Other approaches of self-organized patterning in thin films involve spatially varying patterned destabilizing force fields such as electric field and adhesive forces, etc.[6]

A different strategy for polymer patterning involves formation of surface relief structures in form of wrinkling and buckling of the polymer films under the mechanical stresses generated during stretching-compression, rapid heating-cooling cycles, differential swelling-shrinkage etc.[7] The characteristic length scales of the wrinkling and buckling scales with film thickness and the stresses in the film and substrate.[8]

Another widely studied technique for the patterning of laser-absorbing polymers and other hard materials like metals, glass and ceramics is laser ablation.[9] Depending on its absorption and interaction with the material, laser light may break the atomic bonds or cause material melting and removal or modify it chemically and physically. Use of pulsed lasers (nano, pico or femtosecond) further adds the advantage of increased throughput by making it a continuous fabrication process. One can also use the structured light by incorporating interference and diffraction effects to produce more complex structures.[10,11] Under low laser fluence, swelling or material expansion has been observed as a pre-ablative process, which forms wrinkle like patterns on the surface.[12] However, a low absorption coefficient of the polymers (e.g., poor absorption of 355 nm laser in PS and PMMA[13]) renders melting and subsequent pattern formation difficult unless some polymer specific modifications are made for enhanced optical absorption.[14]

The current work presents the self-organized patterns generated by laser processing of thin (< 400 nm) *non-absorbing* polymer films coated on absorbing substrates. In this technique, the structured light resulting from the near-field diffraction from a small periodic aperture was employed on a thin polymer film coated on silicon. Third harmonic of Nd-YAG laser (355 nm) was used as the incident radiation. A single pulse of laser was sufficient to produce relief structures on the large areas (~cm$^2$) of polymer thin films without any loss of material from the surface, making it a high throughput continuous process. Use of ultrathin polymeric films offers greater flexibility in controlling surface structures via self organization.[6,7] Local melting caused by the absorption of structured light and rapid diffusion of heat in the substrate produce simple to complex structures that could be tuned by changing aperture size and geometry, polymer film thickness and aperture substrate distance. This method is capable of producing ultrafast (~100 ns), cheap, and large area patterns (~cm$^2$) on a variety of polymers as it demands no special optical or chemical properties of the polymer. The patterns thus generated have some interesting and unique characteristics that are clearly differentiated from the patterns formed by other lithography and self-organization based methods: (1) the patterns are in the form of an open lattice-work structure consisting of fine lines with line-widths that are one to two orders of magnitude smaller than the mask openings, (2) the pattern-geometry is far more complex than the mask geometry, (3) both the pattern length scales and morphology can be modulated by changing the film thickness.

## 2. Results and Discussion

Polymer thin films coated on different substrates were irradiated by a single pulse of Nd-YAG laser and the patterns that evolved on the surface of polymer were examined under microscope. **Figure 1** shows the schematic diagram of the experimental setup highlighting different aperture geometries and the process variables: half aperture width, *a* and the aperture substrate distance, $l$. Effects of various process parameters, e.g. aperture size ($a$) and distance ($l$), film thickness, substrate etc., on the evolution of patterns are investigated.

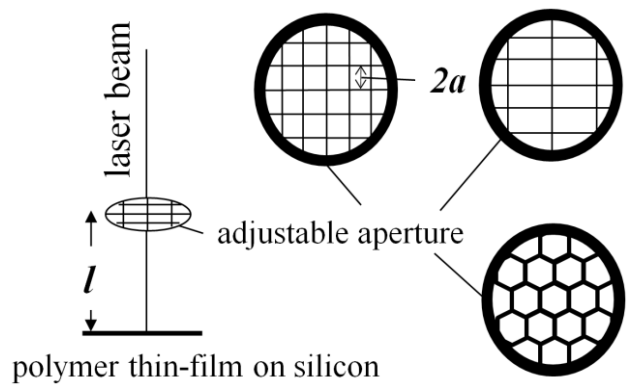

**Figure 1.** Schematic diagram of experimental setup showing adjustable apertures of various types (square, rectangular, hexagonal openings). The distance between film and aperture ($l$) is adjusted by micromanipulator.

Different types of substrates such as silicon, quartz, glass and ITO coated glass were used for coating polymer thin films. With silicon as the substrate for polymeric films, the patterns appeared prominently whereas there were no patterns on the polymer films coated on glass and quartz. Polymer films on ITO coated glass showed some poorly formed pattern. Based on these observations, it is evident that the substrate plays an important role in the proposed method of pattern formation in thin polymer films. For both the polymers (PS and PMMA), absence of any pattern when coated on glass and quartz substrates is due to their transparency to 355 nm light.[13] Silicon, on the other hand, absorbs the radiation energy from the laser pulse and transfers it to the polymer in the form of thermal energy. The local heating and subsequent freezing cause reorganization of the polymer film in the form of wrinkle like surface relief patterns. The laser fluence used in the experiments was not sufficient to cause melting of silicon, which was confirmed by examining the substrate after selective removal of the patterned polymer film. In the subsequent discussion silicon is used as the substrate for the patterning of polymer thin-films.

We first investigated the effect of the laser beam on the polymer film without using any micro-aperture. Randomly distributed wrinkle-like surface relief patterns have appeared on the films of both PS and PMMA. There are two characteristic length scales of these patterns: wrinkle line width ($w_c$) and the mean separation between the wrinkle lines ($L_c$). Both of these are found to increase linearly with the film thickness (**Figure 2**). Relatively thick films (100 nm < $h$ < 400 nm) showed well developed, smooth, but sparse patterns. As the thickness is decreased, the patterns become increasingly denser and fragmented. The line-width of ridges decreased linearly with the film thickness. Figure 2a shows the linear dependence of the mean inter-line distance between the wrinkles, $L_c$ on the film thickness in the case of both the PS and PMMA thin-films. This liner dependence is in the qualitative agreement with the earlier work on the dependence of wrinkle wavelengths on PS films subjected to thermal and mechanical stresses.[8] Based on the refractive index of silicon at 355 nm, the surface reflection loss can be calculated, which is 75.77% at normal incident angle.[15] Further, simple calculations of heat transfer at room temperature showed that the heat conduction through the silicon wafer is extremely fast (within a few hundred nanoseconds).[16] The thermal diffusivity of PS being three orders less (Si: 0.91 cm$^2$ s$^{-1}$, PS: 5.86×10$^{-4}$ cm$^2$ s$^{-1}$), one may expect compressive stresses in the polymer film leading to the formation of ridges on its surface. Use of thin non-absorbing polymer film is an essential factor in this patterning technique because the film should be thin enough to melt and reorganize before the heat diffuses away to the bulk of the conducting substrate. For the laser energies used in these experiments, films up to ~400 nm could be patterned. The variation of the characteristic length-scales of the surface relief structures (**Figure 2**) is useful for optimizing the film thickness for a given diffraction pattern. For example, films thinner than 75 nm are required if a pattern with features of ~ 10 µm is to be fabricated.

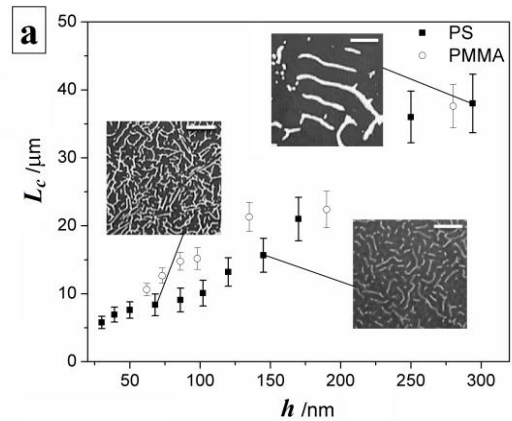

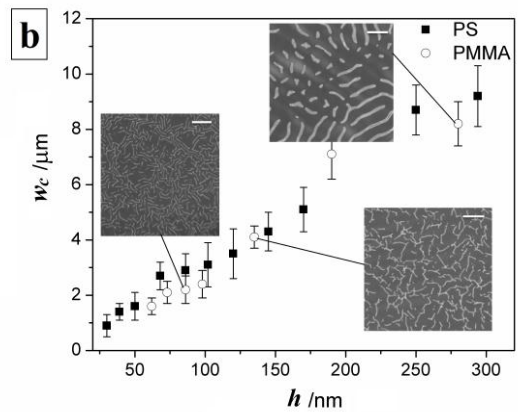

**Figure 2.** Wrinkle patterns formed on PS and PMMA thin films by exposure to single laser pulse without any mask and the dependence of characteristic length-scales of the pattern on the polymer film thickness. (a) Mean separation of wrinkles, $L_c$ as a function of film thickness, $h$ shows linear dependence for both PS and PMMA. Insets show the optical micrographs of the wrinkled PS films of thicknesses 68 nm, 145 nm and 294 nm, respectively. (b) Wrinkle line width, $w_c$ as a function of film thickness, $h$ shows linear dependence for both PS and PMMA. Insets show the optical micrographs of the wrinkled PMMA films of thicknesses 86 nm, 135 nm and 280 nm, respectively. (Scale bar: 50 μm)

Material properties such as molecular weight, thermal conductivity and viscosity are some of the important factors which govern $L_c$ and $w_c$. To decrease the value of $L_c$ and $w_c$, lower molecular weight, higher thermal conductivity and lower melt viscosity are favorable.

Next, we characterize the pattern modification and control by the use of diffraction gratings. The near field diffraction is characterized by the Fresnel Number, $F$ defined as:

$$F = a^2 / \lambda l \qquad (1)$$

Where $a$ is half aperture width, $l$ is distance of aperture from the substrate and $\lambda$ is the wavelength of incident light. Solving near field diffraction equations for a square aperture, one can get intensity distribution on the screen as shown in **Figure 3**.[17] These patterns depend only on $F$, but not on the aperture width ($a$) or screen distance ($l$) individually. Some of the notable features in these patterns are: maxima adjacent to the edges of the square are most prominent, number of intermediate maxima increase with $F$, difference in intensity at bright and dark regions reduces with increasing $F$ and the center appears bright in the case of odd numbered values of $F$. The patterns formed by a square aperture can be thought of as an orthogonal overlap of two identical single slits. By the same argument, one can constitute the near field diffraction patterns of hexagonal aperture as the combination of three identical single slits placed at the 60° angle to each other (simulations for hexagonal as well as rectangular apertures are not shown here).

In the present work, apertures of different size and geometry have been used to establish a better understanding of the effects of diffraction pattern on the relief structures generated on thin polymer films. Results for four different types of apertures are reported here: square apertures of opening sizes 110 µm (aperture $S_1$: $a = 55$ µm) and 36 µm (aperture $S_2$: $a = 18$ µm), a hexagonal aperture of 55 µm (aperture $H_1$: $a = 27.5$ µm) and a rectangular aperture of size 282×36 µm (aperture $R_1$: $a = 18$ µm).

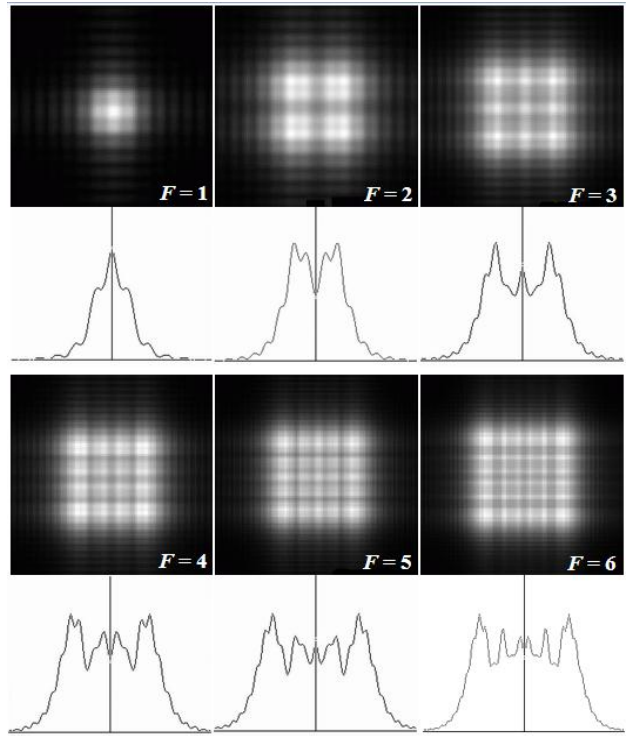

**Figure 3.** 2D Simulated Fresnel diffraction patterns and corresponding 1D intensity profiles for Fresnel numbers, 1 to 6. The relative intensities of radiation maxima and the spatial distribution can be clearly seen.[17]

The correlation between experimentally obtained patterns with the simulated diffraction patterns is illustrated in **Figure 4** for an 81 nm thick PS film with aperture $S_1$. Value of $F$ is calculated for each of the experiment using equation 1. The experimental parameters, $l$ and $a$ are chosen to produce nearly integer values of $F$, so that the diffraction intensity patterns are clearly resolved. For Figure 4a, calculated $F$ is 3.94±0.07 ($a$ = 55 μm, $l$ = 2.16±0.04 mm) while for Figure 4b it is 6.09±0.17 ($a$ = 55 μm, $l$ = 1.40 ± 0.04 mm). In the further discussions, the closest integer value of $F$ is mentioned for the experimentally obtained structures and simulated intensity patterns.

First column images in Figure 4 are of patterned PS film, second column shows simulated intensity pattern corresponding to the same $F$, while third column shows overlap image and draws correspondence of each intensity maxima with pattern on the film. It is evident that the relief structures form where the laser intensity is higher. Interestingly, a unit cell of the polymeric structures guided by the diffraction pattern is already about an order of magnitude smaller than the aperture size, but the line widths are even smaller.

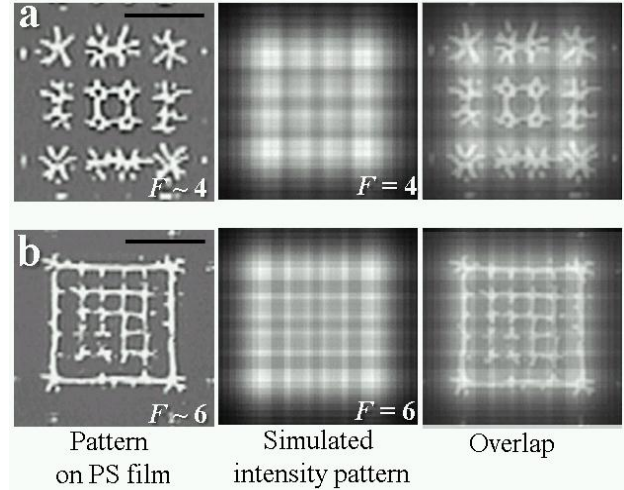

**Figure 4.** Comparison of the optical images of experimentally generated patters on PS film of thickness 81 nm with the simulated images. (a) comparison for $F$~4 ($a$: 55 μm, $l$: 2.16 mm) (b) comparison for $F$~6 ($a$: 55 μm, $l$: 1.40 mm). Overlap image in both the cases show correspondence of intensity maxima with the pattern on the film. (Scale bar: 50 μm).

In **Figure 5**, the effect of film thickness on the pattern formation is illustrated keeping the optical conditions unchanged ($F$, i.e., $a$ and $l$). Figure 5a shows the simulated intensity pattern corresponding to $F$=8. For a 39 nm thick film ($L_c$~7 μm, $w_c$~1.4 μm), the pattern on the polymer film is quite comparable to the simulated image, where patterns appeared corresponding to each intensity maxima (Figure 5b) that are separated by a distance of ~7 μm near the center of the pattern. Therefore, patterning becomes possible on the scale which is about one order smaller than the mask used ($S_1$, opening size 110 μm). With the increased film thickness of 91 nm ($L_c$~9.5 μm, $w_c$~2.8 μm), finer details start to disappear and only the gross pattern with a distorted interior is seen (Figure 5c). For a thicker film (135 nm; $L_c$~ 14 μm, $w_c$~ 4 μm), only the outer square pattern is seen, as the intensity maxima adjacent to the edge is most prominent (Figure 5d). The relief structures corresponding to the weaker intensity maxima completely disappear in thicker films because of the less energy imparted per unit volume of the film. Thus, the size of patterns formed on thin polymer films is determined by a competition between the size of diffraction patterns and the characteristic length scales of the wrinkles ($L_c$ and $w_c$) formed without any aperture.

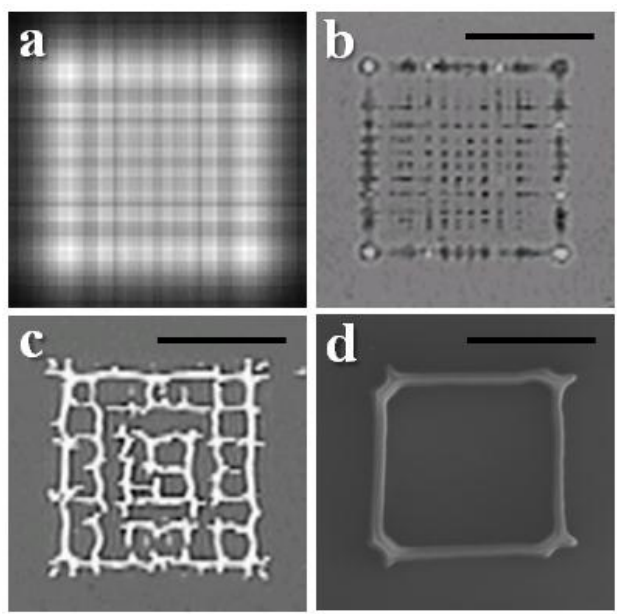

**Figure 5.** Comparison of patterns on PS thin films when the film thickness is varied keeping diffraction conditions constant. (a) Simulated intensity distribution corresponding to $F = 8$. Patterns formed on PS films of thickness (b) 39 nm, (c) 91 nm and (d) 175 nm respectively at $F \sim 8$ (*a*: 55 μm, *l*: 1.04 mm). (Scale bar: 50 μm)

In **Figure 6**, effect of $F$ on the pattern formation is shown on an 81 nm thick PS film ($L_c \sim 9$ μm, $w_c \sim 2.8$ μm) using the aperture $S_1$ (Figure 6a). It is evident from the figure that finer features start appearing within a unit-cell pattern when the aperture is brought closer to the film owing to the finer diffraction patterns at higher $F$ (Figure 3). For relatively small $F$ ($F \leq 4$), the diffraction patterns of neighboring cells interfere to form a more continuous pattern which results in less distinct boundaries of unit-cells. The resulting structure is a connected grid-like structure (Figure 6b, 6c). For higher $F$, the unit-cell pattern acquires a clear identity with a greater density of finer features within each cell. For $F \sim 8$, the inter-line spacing within a box is about 10 μm which is an order of magnitude smaller than the size of opening in the aperture (110 μm) and same as the length scale of the diffraction pattern. The line widths of the polymer pattern are much finer (~ 3 μm). The ordered pattern formation occurs only in that part of the aperture opening where the inter-line spacing ($L_c$ in Figure 2) is less than the distance between the neighboring intensity maxima. For example, $L_c$ for the 81 nm film is about 9 μm and thus ordered formation occurs for all $F$ shown in Figure 6. In contrast, patterns in Figures 5c and 5d replicate the diffraction pattern incompletely owing to $L_c$ of thicker films being larger than the separation between intensity maxima in the imposed diffraction pattern. Also, the gross size of separated unit-cell vary from 65 μm (for $F\sim5$) to 75 μm (for $F\sim8$), which is about 30–40% reduction from the unit-cell size on the aperture. The quality of pattern is not so good in terms of fabrication of high fidelity structures; however, it may provide an effective and quick solution to surface texturing and low precision structures.

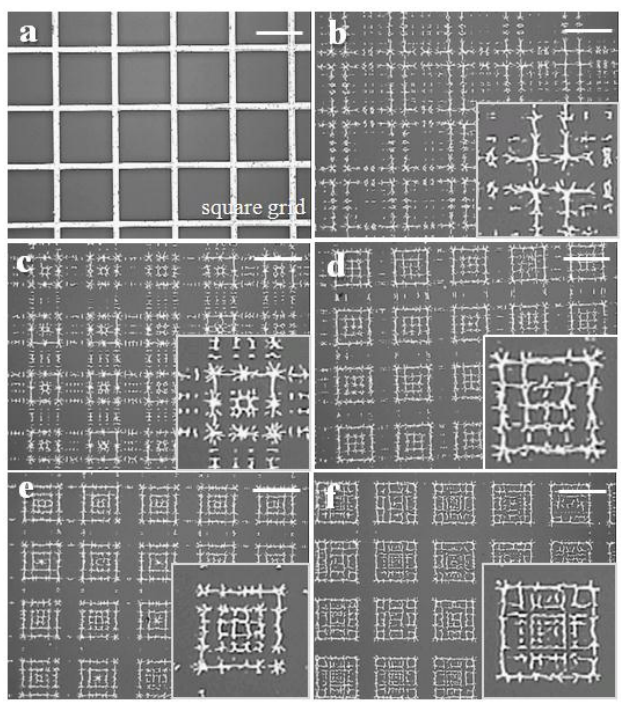

**Figure 6.** Comparison of patterns on PS thin films when diffraction conditions are changed keeping the same film thickness. Optical micrograph (a) shows the square opening TEM grid used for patterning; and (b), (c), (d), (e) and (f) correspond to a patterned PS film of thickness

81 nm and different Fresnel number: $F \sim 2$ (*a*: 55μm, *l*: 4.24 mm), 4 (*a*: 55μm, *l*: 2.12 mm), 5 (*a*: 55μm, *l*: 1.72 mm), 6 (*a*: 55μm, *l*: 1.4 mm) and 8 (*a*: 55μm, *l*: 1.04 mm), respectively. Insets show the enlarged unit-cell pattern formed within each aperture opening. (Scale bar: 100 μm)

Next, patterns generated by a square aperture with smaller openings ($S_2$, $a = 18$ μm) are examined on thin PMMA films. *F*, being a quadratic function of *a*, rapidly drops to lower values. Moreover, for $F \geq 3$, the separation of intensity maxima within a square aperture becomes smaller than the characteristic length scales of the thin-film ($h > 75$ nm) and hence features within unit cell are generally absent. **Figure 7** shows the patterns formed by using aperture $S_2$ (Figure 7a) on 78 and 98 nm thick films. On a film with thickness 98 nm ($L_c \sim 15$ μm, $w_c \sim 2.4$ μm), only one circular ring-like pattern in each aperture opening is observed when the value of *F* is kept nearly 2 (Figure 7b). The diameter of circular pattern is ~18.5 μm with the line width of ~3 μm, both of which are close to the characteristic length scales ($L_c$, $w_c$) of the relief patterns on 98 nm thick PMMA films (Figure 2). Further the size of structure is approximately half of the opening (36 μm) of the aperture used. As the film thickness is decreased to 78 nm ($L_c \sim 13$ μm, $w_c \sim 2.2$ μm), finer features start appearing and pattern becomes increasingly fragmented and multi-branched with the appearance of four-fold symmetric and radially oriented features (Figure 8c and 8d). The overall size of the pattern in this case is about 28 μm, which is ~22% smaller than the mask opening, with the line width of ~2 μm. In such cases, both the ring pattern and the star pattern generated are remarkably robust and well-defined over large areas. Clearly, complex patterns of this nature cannot be produced by photolithography or by any guided self-assembly based technique without using an intricate mask or template of matching complexity.

- 13 -

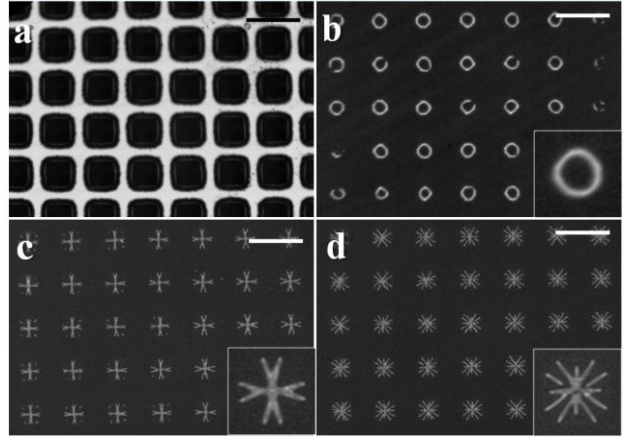

**Figure 7.** Optical micrographs of (a) the 400 mesh square opening TEM grid used for patterning, (b) patterned PMMA film of thickness 98 nm for $F \sim 2$ (*a*: 18 μm, *l*: 0.56 mm), (c) patterned PMMA film of thickness 78 nm $F < 1$ (*a*: 18 μm, *l*: 1.6 mm) and (d) patterned PMMA film of thickness 78 nm $F \sim 2$ (*a*: 18 μm, *l*: 0.48 mm). (Scale bar: 50 μm)

A grid of hexagonal openings ($H_1$, $a = 27.5$ μm) produces six-fold symmetric star shaped patterns on 98 nm thick PMMA films ($L_c \sim 15$ μm, $w_c \sim 2.4$ μm) as shown in **Figure 8**. The gross size of unit-cell pattern is ~36 μm (~35% smaller than the mask opening) with a line width of ~ 2 μm,. Moreover, the six-fold symmetry along with the radial alignment arises from the similar symmetry in the mask. When the aperture is moved away from the film (*F* decreased), a circular pocket appears at the center (images in Figure 8b–d). This may be due to the fact that for lower *F,* laser intensity is concentrated more near the center (Figure 3) and after absorption in the silicon substrate, heat diffuses radially giving rise to the formation of a central circular ring.

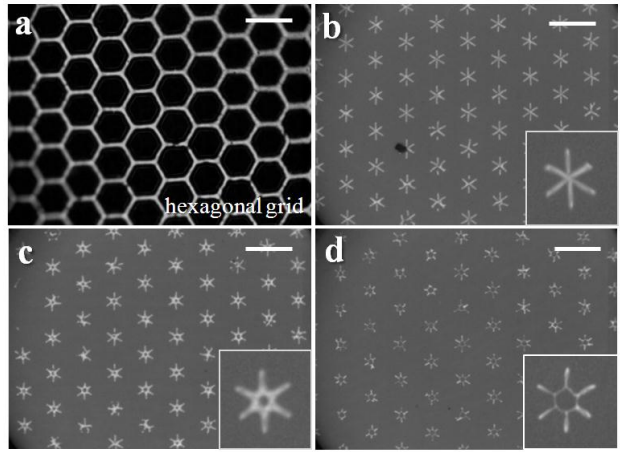

**Figure 8.** Optical micrographs of (a) the hexagonal opening TEM grid used for patterning, (b) $F \sim 9$ ($a$: 27.5 μm, $l$: 0.24 mm), (c) $F \sim 4$ ($a$: 27.5 μm, $l$: 0.52 mm) and (d) $F \sim 1$ ($a$: 27.5 μm, $l$: 1.44 mm) patterned PMMA film of thickness 98 nm. (Scale bar: 100 μm)

Patterns depicted in Figures 7 and 8 are examples of beyond-the-mask patterning where fairly complex shapes that are not otherwise easily fabricated by other means are produced from simple masks.

The patterns created on PS films of thickness 138 and 75 nm by a rectangular aperture ($R_1$, **Figure 9**a) with the opening dimensions of 282×36 μm are shown in Figure 9. Patterns evolved in the form of smooth 2.5 μm thick twin lines separated by ~12 μm (Figure 9b) on a thicker PS film (138 nm, $L_c$~14 μm, $w_c$~4 μm). However, for a thinner film (75 nm, $L_c$~8.5 μm, $w_c$~2.8 μm), patterns were fragmented (Figure 9c and 9d). Moreover, as $F$ is increased from 1 (Figure 9c) to 2 (Figure 9d), splitting of the single line into two parallel lines is observed (the smaller dimension of the rectangular opening is used in calculation of $F$). The width of the pattern (twin or single line) is 12 – 21 μm which is ~ 66% smaller than the mask opening. Further the width of the fine lines bears no relation to the mask opening, but is determined only by the film thickness.

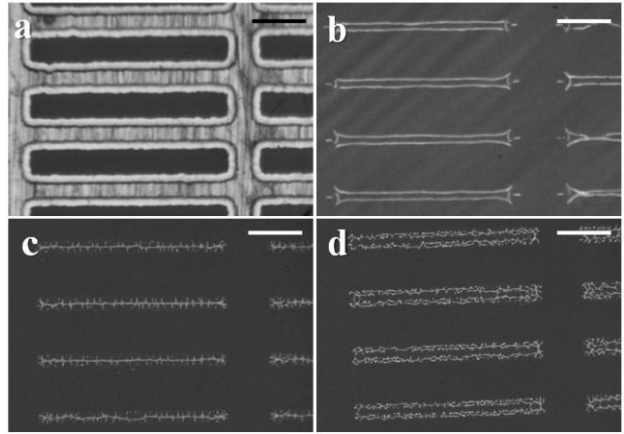

**Figure 9.** Optical micrographs of: (a) rectangular openings of the TEM grid used as a mask, (b) patterned PS film of thickness 138 nm with $F \sim 4$ (*a*: 18 μm, *l*: 0.24 mm), (c) patterned PS film of thickness 75 nm with $F \sim 1$ (*a*: 18 μm, *l*: 0.8 mm) and (d) patterned PS film of thickness 75 nm with $F \sim 2$ (*a*: 18 μm, *l*: 0.48 mm) (Scale bar: 50 μm)

Atomic force microscope surface topographs of these patterns confirm that they are in the form of surface relief structures with the aspect ratio (height to width) of less than 1. **Figure 10a** shows a pattern consisting of square shaped enclosures in a PS film of thickness 175 nm for $F\sim 8$. AFM topograph of a single square shaped enclosure is shown in Figure 10b. It is evident from the height profile that the height of these features is more than the film thickness. In this image, the protrusions are about 300 nm tall formed on a 175 nm thick PS film. Similarly in Figure 10c, star shaped enclosures on a 95 nm thick PMMA film are shown and Figure 10d shows an AFM scan of a single feature from this pattern whose height is about 180 nm.

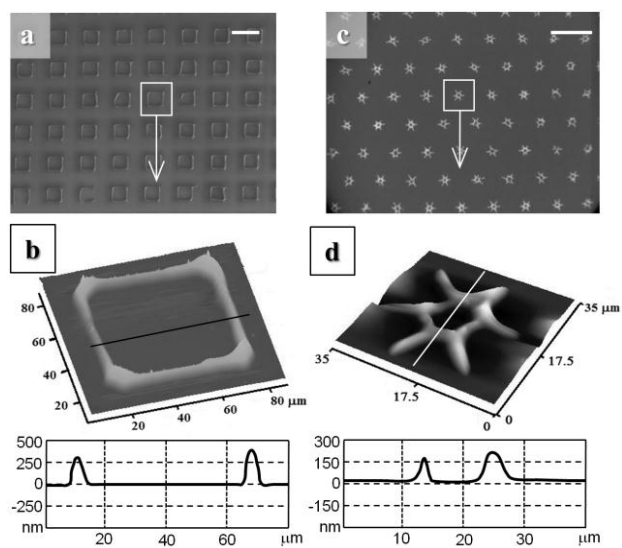

**Figure 10.** (a) SEM micrograph of a patterned PS film of 175 nm thick, (*a*: 55 μm, *l*: 1.04 mm, F~8) (b) AFM topograph of a single feature highlighted in (a), (c) optical micrograph of a patterned PMMA film of 95 nm thick (*a*: 27.5 μm, *l*: 0.52 mm, F~4), (d) AFM topograph of a single feature highlighted in (c). (Scale bars in (a) and (c): 100 μm)

In addition to the high molecular weight non absorbing polymers like PS and PMMA, a smaller organo-metallic molecule, Palladium hexadecylthiolate (PdSH) which absorbs the 355 nm radiation was also employed to assess the role of laser absorption in the patterning media. **Figure 11** compares the patterns generated on thin films of PS (88 nm), PMMA (95 nm) and PdSH (90 nm) using a hexagonal aperture ($H_1$, 55 μm opening). All the patterns were made using identical optical conditions, i.e. $F \sim 9$ (*a*: 27.5 μm, *l*: 0.24 mm). It is evident from the images that the response to the laser pulse is qualitatively same in the case of the non-absorbing polymers (PS and PMMA) which shows only a few broad patterns. On the other hand PdSH which absorbs the radiation shows entirely different pattern with all the intricate features of the diffraction pattern reproduced inside. It becomes possible by the low melt viscosity of lower molecular wt. PdSH which gives faster kinetic response in the structure formation on thin-films.

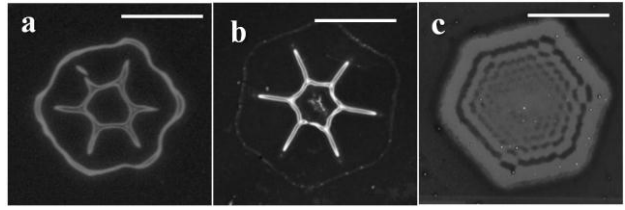

**Figure 11.** Patterns produced with hexagonal aperture on (a) PS film (88 nm thick), (b) PMMA film (95 nm) and (c) PdSH film (90 nm thick) using identical optical conditions ($F$~9, $a$: 27.5 μm, $l$: 0.24 mm). (Scale bar: 20 μm)

In principle one could keep on decreasing the aperture size and aperture distance to produce smaller and smaller patterns. But there are certain constrains that need to be addressed to make submicrometer features viable. To achieve submicrometer benchmark it is important that the both Lc and wc are in sub-micrometer range, which may be possible for small molecules or low molecular weight polymers. Moreover, to bring in near field diffraction effects ($F > 1$) with smaller aperture width ($a$) both laser wavelength (λ) and aperture distance ($l$) should be as small as possible (Equation 1). Laser wavelength cannot be decreased much below 200 nm (deep UV laser); therefore, for a 2 μm opening aperture we need to work with the aperture distance close to 1 μm to achieve $F = 5$, which is only possible by high precision nano-manipulators. Moreover, fragmentation of patterns are observed for lower $L_c$ with or without using any aperture; probably, because of the lower inertia of thinner films to prevent chaotic wrinkling. Smoothness of the grid bar is also very critical in the formation of sharp diffraction pattern. For smaller grids the edge roughness of the grid is even more critical.

## 3. Conclusions

An ultrafast large area micro-patterning technique of thin-films of non-absorbing polymers (PS and PMMA) using nanosecond pulsed laser is demonstrated, which proves to be a simple, flexible and direct method. Exposure to a single laser pulse (~8 ns) causes absorption in form of heat in silicon which provides sufficient energy to the polymer film to locally melt, reorganize and rapidly freeze to form wrinkle like surface relief structures. Characteristic length scales ($L_c$, $w_c$) of relief structures formed on laser exposure without aperture are found to increase linearly with the film thickness. Near field diffraction effects in the patterning are incorporated by placing small periodic apertures of different sizes and geometry and are fine-tuned by changing the distance between aperture and thin-film. Relief patterns on the thin-films are formed by a competition between the characteristic length scales of wrinkle formation and the imposed diffraction pattern. Using a large opening aperture with large value of $F$ on thinner polymer films (< 50 nm), one order smaller patterns than the mask are obtained, however, the fidelity of these structures is poor. Relatively thicker films (>90 nm) produce smooth patterns of some regular shapes such as square, circular, star shaped patterns and open micro-channels, etc. that can find applications in microfluidics and biological micro-arrays. Further, multi-fold symmetric complex structures are observed in thinner films (<90 nm) with smaller apertures. Also, a size reduction of 22 – 66% is observed in the unit-cell of the patterns formed as compared to the aperture opening. Finally, an absorbing film of PdSH is shown to produce much complex structure aided by low viscosity hence better kinetics of the material to form these structures.

## 4. Experimental Section

Nd-YAG nanosecond pulsed laser (Quanta-Ray® Lab, Spectra Physics) is employed as the coherent light source for the experiments. 355 nm wavelength UV radiation is used which is the third harmonic of its fundamental wavelength (1064 nm). The fluence is varied between 100–400 mJ cm$^{-2}$ and tuned around the onset of visible surface patterns avoiding film rupture, which works out to be 250 mJ cm$^{-2}$. Unless otherwise mentioned, in all the experiments only a single laser pulse (~8 ns) of 250 mJ cm$^{-2}$ fluence was used. All the experiments were done with parallel laser beam without using any kind of lens or focusing mechanism. Experimental setup, as shown in Figure 1, consists of a periodic aperture, i.e., commercially available TEM grids namely Gilder 200 mesh grids, Veco 400 mesh grids, Veco hexagonal 300 mesh grids and Veco Slotted Pattern 75/300 mesh grids. Gilder 200 mesh grids have square aperture of 110 μm, Veco 400 mesh grids have square aperture of 36 μm, Veco hexagonal 300 mesh grids have 55 μm openings, while Veco Slotted Pattern 75/300 mesh grids have rectangular opening of 282×36 μm. The aperture is held using a micromanipulator to allow changing the distance between aperture and the substrate precisely (precision x-y: 100 μm, z: 40 μm). PS and PMMA of average molecular weights, 280,000 and 120,000 respectively were procured from Sigma Aldrich. Thin-films of thickness ranging from 30 nm to 400 nm were spin coated on thoroughly cleaned substrates (silicon wafer/quartz/glass) by using 0.5–5 w/v % polymer solutions in HPLC grade toluene. PdSH, $Pd(SC_{16}H_{33})_2$ was prepared by mixing and vigorously stirring equal moles of Pd acetate and hexadecanethiol in toluene.[18] The resulting thiolate was dried and washed with methanol and acetonitrile to remove excess thiol and then dissolved in toluene. Thin films of this material were spin coated on cleaned substrates using this solution. Thickness measurements of thin films coated on substrates were carried out using ellipsometer (Nanofilm, $EP^3$-SE). Imaging of the patterned films was done using optical microscope (Leica DMLB 100s), field emission scanning electron microscope (Zeiss Supra 40VP) and atomic force microscope (Veeco NanoMan). A java applet is used to calculate the diffraction pattern and intensity distribution for different Fresnel numbers.[17] Calculation of mean separation between wrinkle lines of random surface relief patterns generated from the exposure of polymer films with laser without any mask, is done by drawing 6×8 grid in three different images and taking the mean value of distance between two intersections and standard deviation as the error bars. Line width of these patterns was calculated by taking 10 point average in three different images.

**Acknowledgements**

This work is supported by a DST IRHPA grant and Units of Nanosciences at IIT Kanpur and JNCASR Bangalore. A.V. acknowledges the initial efforts of Sruthi Murlidharan in setting up the experiments. Useful discussions regarding heat transfer calculations in silicon and polymer film with Ramki Kalyanaraman and Jeremy Strader are gratefully acknowledged.

[1] a) T. R. Hebner, C. C. Wu, D. Marcy, M. H. Lu, J. C. Sturm, *Appl. Phys. Lett.* **1998**, 72, 519; b) Z. Nie, E. Kumacheva, *Nature Mater.* **2008**, 7, 277.

[2] a) T. B. Cao, Q. B. Xu, A. Winkleman, G.M. Whitesides, *Small* **2005**, 1, 1191; b) P. Greil, *Adv. Eng. Mater.* **2000**, 2, 339.

[3] a) M. Geissler, Y. Xia, *Adv. Mater.* **2004**, 16, 1249; b) Y. Xia, G. M. Whitesides, *Angew. Chem. Int. Ed.* **1998**, 37, 550.

[4] a) A. M. Higgins and R. A. L. Jones, *Nature* **2000**, *404*, 476; b) A. Sehgal, V. Ferreiro, J. F. Douglas, E. J. Amis, A. Karim, *Langmuir* **2002**, *18*, 7041; c) K. Y. Suh, H. H. Lee, *Adv. Mater* **2002**, *14*, 346; d) D. Julthongpiput, W. Zhang, J. F. Douglas, A. Karim, M. J. Fasolka, *Soft Matter* **2007**, *3*, 613; e) M. Geoghegan, C. Wang, N. Rhese, R. Magerle and G. Krausch, *J. Phys: Condens. Matter* **2005**, *17*, S389; f) M. Cavallini, C. Albonetti, F. Biscarini, *Adv. Mater.* **2009**, *21*, 1043; g) B. Yoon, H. Acharya, G. Lee, H. -C. Kim, J. Huh, C. Park, *Soft Matter* **2008**, *4*, 1467; h) G. G. Baralia, C. Filiatre, B. Nysten and A. M. Jonas, *Adv. Mater.* **2007**, *19*, 4453; i) D. H. Kim, M. J. Kim, J. -Y. Park, H. H. Lee, *Adv. Funct. Mater* **2005**, *15*, 1445.

[5] a) E. Schaffer, T. Thurn-Albrecht, T. P. Russell, U. Steiner, *Nature* **2000**, *403*, 874; b) E. Schaffer, T. Thurn-Albrecht, T. P. Russell, U. Steiner, *Europhys. Lett.* **2001**, *53*, 518; c) S. Srivastava, D. Bandyopadhyay, A. Sharma, *Langmuir* **2010**, *26*, 10943; d) S. Srivastava, P. D. S. Reddy, C. Wang, D. Bandyopadhyay, A. Sharma, *J. Chem. Phys.* **2010**, *132*, 174703; e) S.

Y. Chou, C. Keimel, J. Gu, *Nature* **2002**, *417*, 835; f) M. Gonuguntla, A. Sharma, S. A. Subramanian, *Macromolecules* **2006**, *39*, 3365.

[6] a) G. Reiter, *Phys. Rev. Lett.* **1992**, *68*, 75; b) G. Reiter, *Langmuir* **1993**, *9*, 1344; c) A. Sharma, G. Reiter, *J. Colloid Interface Sci.* **1996**, *178*, 383; d) R. Xie, A. Karim, J. F. Douglas, C. C. Han, R. A. Weiss, *Phys. Rev. Lett.* **1998**, *81*, 1251; e) R. V. Craster, O. K. Matar, *Rev. Mod. Phys.* **2009**, *81*, 1131.

[7] a) J. S. Sharp, R. A. L. Jones, *Adv. Mater.* **2002**, *14*, 799; b) P. J. Yoo, K. Y. Suh, S. Y. Park, H. H. Lee, *Adv. Mater.* **2002**, *14*, 1383; c) E. P. Chan, A. J. Crosby, *Adv. Mater.* **2006**, *18*, 3238; d) A. Schweikart, A. Fery, *Microchim Acta* **2009**, *165*, 249; e) J. Y. Chung, A. J. Nolte, C. M. Stafford, *Adv. Mater.* **2009**, *21*, 1358; f) J. S. Sharp, K. R. Thomas, M. P. Weir, *Phys. Rev. E* **2007**, *75*, 011601; g) X. He, J. Winkel, W. T. S. Huck, *Adv. Mater.* **2009**, *21*, 2083; h) J. Genzer, J. Groenewold, *Soft Matter* **2006**, *2*, 310; i) E. Cerda, L. Mahadevan, *Phys. Rev. Lett.* **2003**, *90*, 074302; j) K. Efimenko, M. Rackaitis, E. Manias, A. Vaziri, L. Mahadevan, J. Genzer, *Nat. Mater.* **2005**, *4*, 293; k) C. -C. Lin, F. Yang, S. Lee, *Langmuir* **2008**, *24*, 13627.

[8] a) R. Huang, *J. Mech. Phys. Solids* **2005**, 53, 63; b) J. Groenewald, *Physica A* **2001**, 298, 32; c) H. Huang, J. Y. Chung, A. J. Nolte, C. M. Stafford, *Chem. Mater.* **2007**, 19, 6559; d) J. Y. Chung, T. Q. Chastek, M. J. Fasolka, H. W. Ro, C. M. Stafford, *ACS Nano* **2009**, 3, 844.

[9] a) T. Lippert, J. T. Dickinson, *Chem. Rev.* **2003**, 103, 453; b) R. Braun, R. Nowak, P. Hess, H. Oetzmann, C. Schmidt, *Appl. Surf. Sci.* **1989**, 43, 352; c) D. Mills, K. W. Kolasinski, *J. Phys. D: Appl. Phys.* **2005**, 38, 632; d) B. K. Nayak, M. C. Gupta, K. W. Kolasinski, *Nanotechnology* **2007**, 18, 195302; e) N. S. Murthy, R. D. Prabhu, J. J. Martin, L. Zhou, R. L. Headrick, *J. Appl. Phys.* **2006**, 100, 023538; f) P. E. Dyer, M. Pervolaraki, T. Lippert, *Appl. Phys. A* **2005**, 80, 529; g) G. B. Blanchet, Yueh-Lin Loo, J. A. Rogers, F. Gao, C. R. Fincher, *Appl. Phys. Lett.* **2003**, 82, 463; h) W. Pfleging, M. Torge, M. Bruns, V. Trouillet, A. Welle, S.

Wilson, *Appl. Surf. Sci.* **2009**, 255, 5453; i) J. Trice, C. Favazza, D. Thomas, H. Garcia, R. Kalyanaraman, R. Sureshkumar, *Phys. Rev. Lett.* **2008**, 101, 017802;

[10] a) A. Lasagni, M. Cornejo, F. Lasgani, F. Muecklich, *Adv. Eng. Mater.* **2008**, 10, 488; b) S. Shoji, Hong-Bo Sun, S. Kawata, *Appl. Phys. Lett.* **2003**, 83, 608; c) T. Vijaykumar, N. S. John, G. U. Kulkarni, *Solid State Sc.* **2005**, 7, 1475; d) M. Y. Shen, C. H. Crouch, J. E. Carey, R. Younkin, E. Mazur, M. Sheehy, C. M. Friend, *Appl. Phys. Lett.* **2003**, 82, 1715; e) E. Haro-Poniatowski, E. Fort, J. P. Lacharme, C. Ricolleau, *Appl. Phys. Lett.* **2005**, 87, 143103; f) S. Jeon, V. Malyarchuk, J. O. White, J. A. Rogers, *Nano Lett.* **2005**, 5, 1351.

[11] a) P. E. Dyer, J. Mackay, C. D. Walton, *Optics Comm.* **2004**, 240, 391; b) P. E. Dyer, S. M. Maswadi, C. D. Walton, M. Ersoz, P. D. I. Fletcher, V. N. Paunov, *Appl. Phys. A* **2003**, 77, 391; c) J. A. Rodrigo, T. Alieva, M. L. Calvo, *Optics Express* **2009**, 17, 4976;

[12] A. F. Lasagni, D. F. Acevedo, C. A. Barbero, F. Mucklich, *Polym. Eng. Sci.* **2008**, 2367.

[13] a) T. Inagaki, E. T. Arakawa, R. N. Hamm, M. W. Williams, *Phys. Rev. B* **1977**, 15, 3243; b) T. Li, C. Zhou, M. Jiang, *Polymer Bulletin* **1991**, 25, 211.

[14] F. Beinhorn, J. Ihlemann, K. Luther, J. Troe, *Appl. Phys. A* **1999**, 68, 709.

[15] a) M. Born, E. Wolf, *Principles of Optics*, 7[th] *ed.*, Cambridge University Press, **1999**, Chap. 14; b) M. A. Green, M. J. Keevers, *Prog. in Photovoltaics Res. and Appl.* **1995**, 3, 189.

[16] R. Kalyanaraman (private communication).

[17] N. Betancort, Fresnel Diffraction Applet, Digital Library Network for Engineering and Technology, Virginia Tech, http://www.dlnet.vt.edu/.

[18] P. J. Thomas, A. Lavanya, V. Sabareesh, G. U. Kulkarni, *Proc. Indian Acad. Sci. Chem. Sci.* **2001**, 113, 611.